\documentclass[12pt]{article}
\usepackage{amsthm}
\usepackage{amsmath}
\usepackage{amsfonts}
\usepackage{amssymb}
\usepackage{amscd}
\usepackage{enumerate}
\usepackage[automark]{scrpage2}                 

\usepackage{dsfont}
\usepackage{amscd}
\usepackage{cite}

\def\A{\mathcal{A}}

\def\R{\mathbb{R}}
\def\C{\mathbb{C}}
\def\H{\mathcal{H}}

\def\p{\psi}

\def\2{\frac{1}{2}}

\def\Tr{\mathrm{Tr}}

\def\be{\begin{equation}}
\def\ee{\end{equation}}
\def\bp{\begin{proof}}
\def\ep{\end{proof}}
\def\bc{\begin{cases}}
\def\ec{\end{cases}}

\newcommand{\bra}[1]{\ensuremath{\left\langle #1\right|}}
\newcommand{\ket}[1]{\ensuremath{\left|#1\right\rangle}}
\newcommand{\braket}[2]{\ensuremath{\left\langle #1\vphantom{#2}\right.\left|\vphantom{#1}#2\right\rangle}}

\newcommand{\ev}[1]{\ensuremath{\left\langle #1\right\rangle}}

\numberwithin{equation}{section}

\begin{document}
\title{Geometrical Description of Quantum Mechanics - Transformations and Dynamics}
\author{G. Marmo$^{1,2}$, G. F. Volkert$^{1,3}$}
\date{ }
\maketitle
{\emph {\footnotesize $^1$Dipartimento di Scienze Fisiche dell'Universit\`a di Napoli Federico II and INFN, Sezione di Napoli,  Complesso Universitario di Monte S. Angelo, via Cintia,\\ I-80126 Napoli, Italy }}\\
{\emph {\footnotesize $^2$ MECENAS, Universit\`a di Napoli Federico II and Universit\`a di Bari, Italy}}\\
{\emph {\footnotesize $^3$ Mathematisches Institut, Ludwigs-Maximilians-Universit\"at, 
Theresienstr. 39,\\ D-80333 M\"unchen, Germany}\\\\
{\footnotesize E-mail: marmo@na.infn.it, volkert@mathematik.uni-muenchen.de}}

\begin{abstract}
In this paper we review a proposed geometrical formulation of quantum mechanics. We argue that this geometrization makes available mathematical methods from classical mechanics to the quantum frame work. We apply this formulation to the study of separability and entanglement for states of composite quantum systems.   
\end{abstract}

\section{Introduction}
There are several sound reasons to try to formulate quantum mechanics in geometrical terms. For instance the high degree of geometrization of classical mechanics, general relativity, gauge theories and others acts as a stimulus to geometrize quantum theories to better understand the quantum-classical transition, to formulate quantum gravity, and to better understand which structures usually dealt with in quantum theories should be attributed to the `system' and which to the `measuring apparatus' (observer) according to the `Heisenberg cut'. In the spirit of Einstein's minimal assumptions, the geometrization would bring out those algebraic structures which should be `dynamically determined', i. e., obtained as solution of the Einstein equations when the distribution of energy and matter in the universe is given.  Thus, by geometrization of quantum mechanics we mean to replace the used description on the Hilbert space by a description on Hilbert manifolds. In this respect, the proposal is very much similar to the transition from special relativity to general relativity: Space-time is considered to be a Lorentzian manifold and the properties of the Minkowski space time are transferred to the tangent space at each point of the space-time manifold. In particular, we go from the scalar product $\eta_{\mu\nu}X^{\mu}X^{\nu}$ to the Lorentzian metric tensor field $\eta_{\mu\nu}dx^{\mu}\otimes dx^{\nu}$,  which is further generalized to non-flat space-time manifolds in the form   $\eta_{\mu\nu}\theta^{\mu}\otimes\theta^{\nu}$ where $\{\theta^{\mu}\}$ are general 1-forms which carry the information on the non-vanishing of the curvature tensor.\\
Similarly, in the geometrization of quantum mechanics we go from the scalar product $\braket{\p}{\p}$ on the Hilbert space to the Hermitian tensor field on the Hilbert manifold, written as $\braket{d\p}{d\p}$.   This would be the associated covariant (0,2)-tensor field.\\
If we consider as starting carrier space not $\H$ but its dual $\H^*$, say not ket-vectors but bra-vectors, in Dirac's notations, we would obtain a (2,0)-tensor field, i.e., a contra-variant tensor field. Once we consider these replacements, algebraic structures will be associated with tensorial structures and we have to take into account that there will be no more invertible linear transformations but just diffeomorphisms. 
The linear structure will emerge only at the level of the tangent space and will `reappear' on the manifold carrier space as a choice of each observer, according to the `Heisenberg cut' \cite{Ercolesi:2007}. We must stress that `manifold descriptions' are naturally appearing already in the standard approach by means of Hilbert spaces when, due to the probabilistic interpretation of quantum mechanics, we realize that pure states are not vectors in $\H$ but rather equivalence classes of vectors, i.e., rays. The set of rays, say $\mathcal{R}(\H)$, is the complex projective space associated with $\H$, it is not linear and carries a manifold structure with `model space' the tangent space at each point $[\p]$. This space may be identified with the Hilbert subspace of vectors orthogonal to $\psi$.\\
In geometrical terms, the transition from non-zero vectors in $\H$ to the corresponding rays defines a principal  $\C_0$-bundle on the total space $\H_0$, with base space $\mathcal{R}(\H)$. By $\C_0$ and $\H_0$ we mean $\C $ and $\H$ respectively without the zero element.\\ 
Other examples of `natural manifolds' are provided by the set of density states which do not allow for linear combinations but only convex combinations. They contain submanifolds of density states with fixed rank. Of course, the group of unitary transformations provides us with another manifold (group-manifold) whose model space, the tangent space at each point, is simply the Lie algebra of anti-Hermitian operators.\\
The best known example of a manifold of quantum states is provided by the coherent states or any generalized variant \cite{Perelomov:1971bd,Gilmore:1972,Onofri:1974}, including also non-linear coherent states \cite{Man'ko:1996xv}.  As it is well known, these manifolds of quantum states, allow to describe many properties of the system we are considering by means of finite dimensional smooth manifolds. Our approach will consider various tensor fields on these manifolds which allow to describe observables, states along with separability and entanglement when we deal with composite systems.\\
As it is well known, most operators of physical interest are unbounded, therefore their description at the manifold level will require to consider domain problems which obscure the geometrical picture. To avoid these technical problems we restrict all our considerations of general character to finite dimensional Hilbert spaces, i.e., quantum systems with a finite number of energy levels. Once the geometrical structures are established one may go to the more realistic situation of infinite dimensional Hilbert spaces and tackle various aspects. 
Some papers on the geometric formulation of quantum mechanics are available. We give here a partial list of these papers we are aware of in the references \cite{Heslot:1985,Rowe:1980,Cantoni:1975,Cantoni:1977a,Cantoni:1977b,Cantoni:1980,Cantoni:1985,Cirelli:1983,Cirelli:1984,Abbati:1984,Ashtekar:1997ud,Gibbson:1992,Brody:2001,deGosson:2007,Man'ko:2005:gqm,Carinena:2007:gqm,Clemente:2008}.  
 
\section{From Hilbert spaces to Hilbert manifolds}    
Here we would like to consider more closely how to replace a vector space with a manifold. To avoid technicalities we shall restrict our considerations to finite vector dimensional spaces (For the manifold point of view for infinite dimensional vector spaces see \cite{Chernoff:1974,Schmid:1987,Lang:1994}). Moreover our manifolds will always be real manifolds so they carry the usual differential calculus. In simple terms this means that we consider the differential calculus on complex-valued functions depending on real variables. A manifold is characterized saying that each point has a neighborhood diffeomorphic to an open subset of its tangent space at that point. For instance, the simplest vector space $\R$ is diffeomorphic to the open interval $(-1,1)$. From the manifold point of view the two sets are equivalent. For both of them the tangent space at each point is $\R$.\\
In finite dimensions any vector space $V$ is isomorphic (although in a basis dependent way, i.e., not naturally isomorphic) with its dual space, the space of scalar valued linear maps on $V$. In particular, for a Hilbert space we may consider as a `starting vector space' either the vector space of kets or the vector space of bras, to use Dirac's notations. If we introduce an orthonormal basis $\{\ket{e_j}\}_{j\in J}$ for $\H$, we define coordinate functions by setting
\be \braket{e_j}{\p} = z^j(\psi),\ee 
usually written simply as $z^j$. By using the dual basis $\{\bra{e_j}\}$ we find 
\be  \braket{\p}{e_j}= \bar{z}_j(\p^*),\ee
this means that coordinate functions $\{z^j\}_{j\in J}$ are defined on $\H$, while coordinate functions $\{ \bar{z}_j\}_{j\in J}$ are defined on the dual space $\H^*$. By using the inner product we can identify $\H$ and $\H^*$. This provides two possibilities: The scalar product $\braket{\p}{\p}$ gives rise to a covariant Hermitian (2, 0)-metric tensor on $\H$
\be \braket{d\p}{d\p} =\sum_j \braket{d\p}{e_j}\braket{e_j}{d\p}= d\bar{z}_j\otimes dz^j,\ee
where we have used $d\braket{e_j}{\p}=\braket{e_j}{d\p}$, i.e., the chosen basis is not `varied',
or to a contra-variant (0,2) tensor 
\be \braket{\frac{\partial}{\partial \p}}{\frac{\partial}{\partial \p}} = \frac{\partial}{\partial \bar{z}_j}\otimes \frac{\partial}{\partial z^j}\ee
on $\H^*$.
\\\\
\emph{Remark:} By considering a changing basis, a `moving frame', we should deal with covariant differential calculus.\\\\
By introducing real coordinates, say 
\be z^j({\p}) = x^j(\p)+iy^j(\p)\ee 
we find       
\begin{eqnarray}
 \braket{d\p}{d\p} =& (dx_j -idy_j)\otimes (dx^j +idy^j)
\\=& (dx_j \otimes dx^j + dy_j\otimes dy^j)+i(dx_j\otimes dy^j - dy_j\otimes dx^j).  
\end{eqnarray}
This expression shows very clearly that the Hermitian tensor is equivalent to a symmetric Euclidean metric tensor (more generally a Riemannian tensor) and a skew-symmetric tensor (a symplectic 2-form).\\
Similarly, on $\H^*$ we may consider    
\begin{eqnarray}
 \braket{\frac{\partial}{\partial \p}}{\frac{\partial}{\partial \p}}=& \bigg( \frac{\partial}{\partial x_j}+i\frac{\partial}{\partial y_j}\bigg)\otimes  \bigg( \frac{\partial}{\partial x^j}-i\frac{\partial}{\partial y^j}\bigg)
\\
=& \bigg(\frac{\partial}{\partial x_j} \otimes \frac{\partial}{\partial x^j} + \frac{\partial}{\partial y_j} \otimes \frac{\partial}{\partial y^j}\bigg)+i\bigg( \frac{\partial}{\partial y_j} \otimes \frac{\partial}{\partial x^j}-\frac{\partial}{\partial x_j} \otimes \frac{\partial}{\partial y^j}\bigg).  
\end{eqnarray}
This tensor field, in contravariant form, may be also considered as a bi-differential operator, i.e., we may define a binary bilinear product on real smooth functions by setting 
\be ((f, g)) = \bigg( \frac{\partial f}{\partial x_j}+i\frac{\partial f}{\partial y_j}\bigg)\cdot  \bigg( \frac{\partial g}{\partial x^j}-i\frac{\partial g}{\partial y^j}\bigg)\ee
which decomposes into a symmetric bracket 
\be (f, g) = \frac{\partial f}{\partial x_j}\frac{\partial g}{\partial x^j} + \frac{\partial f}{\partial y_j}\frac{\partial g}{\partial y^j}\ee
and a skew-symmetric bracket
\be \{f, g\} = \frac{\partial f}{\partial y_j}\frac{\partial g}{\partial x^j} - \frac{\partial f}{\partial x_j}\frac{\partial g}{\partial y^j}.\ee
This last bracket defines a Poisson bracket on smooth functions defined on $\H$.\\
Summarizing, we can replace our original Hilbert space with an Hilbert manifold, i.e. an even dimensional real manifold on which we have tensor fields in covariant form 
\be g = dx_j\otimes dx^j + dy_j\otimes dy^j\ee 
\be \omega = dy_j\otimes dx^j -  dx_j\otimes dy^j,\ee
or tensor fields in contravariant form
\be G = \frac{\partial}{\partial x_j} \otimes \frac{\partial}{\partial x^j} + \frac{\partial}{\partial y_j} \otimes \frac{\partial}{\partial y^j}\ee
\be \Lambda = \frac{\partial}{\partial y_j} \otimes \frac{\partial}{\partial x^j}-\frac{\partial}{\partial x_j} \otimes \frac{\partial}{\partial y^j},\ee 
along with a complex structure tensor field
\be J = dx^j\otimes \frac{\partial}{\partial y_j}-dy^j\otimes \frac{\partial}{\partial x^j}.\ee 
The contravariant tensor fields, considered as bi-differential operators define a symmetric product and a skew symmetric product on real smooth functions. The skew-symmetric product actually defines a Poisson bracket. Once the manifold point of view has been selected (say, $\R$ has been replaced with (-1,1)) we have no meaning for linear transformations, now only diffeomorphisms are available.\\
To recover unitary transformations, we restrict to diffeomorphisms which preserve the Poisson bracket (they are canonical transformations) and moreover preserve the symmetric product (they are isometries for the metric tensor). Their generators at the infinitesimal level are Hamiltonian vector fields, which are also Killing vector fields. It is not difficult to show that `Hamiltonian functions' which define vector fields satisfying previous requirements are necessarily quadratic functions associated with Hermitian matrices. Indeed, the group of unitary diffeomorphisms emerges as the intersection of the group of canonical transformations with the group of isometries.\\
As a further bonus, the symmetric bracket, when restricted to these particular quadratic functions defines a Jordan algebra which is compatible with the commutator bracket so that they define a Lie-Jordan algebra. By restricting our bracket $((f, g))$ to functions whose real and imaginary parts are made of these quadratic functions, we have a new product 
\be f \ast g = ((f, g)).\ee 
This product is associative and compatible with the complex conjugation. By introducing a norm by means of the symmetric product, we obtain a $\C^*$-algebra. What should be stressed is that we have defined unitary diffeomorphisms and $\C^*$-algebras by using only structures available on a real smooth manifold of even dimensions. The original Hilbert space was instrumental to define the tensor fields but we have not used the vector space structure any more.\\
As it is well known, due to the probabilistic interpretation of quantum mechanics, states of a quantum system are to be identified with rays of the Hilbert space according to the equivalence relation 
\be \p_1 \sim \p_2 :\Leftrightarrow \exists \lambda \in \C_0: \p_1=\lambda\p_2\ee
on any two vectors $\psi_1, \p_2 \in\H_0$. If we restrict our attention to the space of rays, $\mathcal{R}(\H)$, we deal with a manifold which is no more diffeomorphic to a vector space.\\ 
Before continuing with general structures and arguments, let us consider the most simple non-trivial example $\H=\C^2$, the Hilbert space of a two-level system. We have 
\be \ket{e_1} = \left(\begin{array}{c}1 \\0\end{array}\right),\,\,\,\,\,\,\ket{e_2} = \left(\begin{array}{c}0 \\1\end{array}\right)\ee
\be z^1=x^1+iy^1,\,\,\,\,\,\, z^2=x^2+iy^2.\ee  
A generic Hermitian matrix $A$ may be decomposed by means of the Pauli-matrices into
\be A = Y_0\sigma_0  + \vec{Y}\vec{\sigma}.\ee 
A generic quadratic form whose Hamiltonian vector field is also Killing is given by 
\be f_A(\p) = \bra{\p}A\ket{\p}= (\bar{z}_1,\bar{z}_2) A \left(\begin{array}{c}z_1 \\z_2\end{array}\right).\ee
Complex valued functions which define the $\C^*$-algebra have the form 
\be F= f_A+if_B = f_{A+iB}\ee 
with an associative product
\be F\ast G =   f_{A+iB}\ast  g_{M+iN} =  f_{(A+iB)(M+iN)}\ee
This resulting product is not pointwise, i.e., it is a non-local product, which is an essential ingredient to take into account the quantum nature of the system we are describing.\\
Let us now describe the manifolds of rays. The most efficient way is to consider coordinate functions, say $(z_1, z_2)$, and consider pure states as rank-one projectors
\be \ket{\p}\bra{\p}= \left(\begin{array}{c}z_1 \\z_2\end{array}\right)(\bar{z}_1,\bar{z}_2)= \left(\begin{array}{cc}z_1\bar{z}_1 & z_1\bar{z}_2 \\z_2\bar{z}_1 & z_2\bar{z}_2\end{array}\right).\ee  
To normalize it we set
\be \rho_{\p} =\frac{1}{z_1\bar{z}_1 + z_1\bar{z}_2}\left(\begin{array}{cc}z_1\bar{z}_1 & z_1\bar{z}_2 \\z_2\bar{z}_1 & z_2\bar{z}_2\end{array}\right).\ee
We notice that 
\be \rho_{\p}\cdot \rho_{\p}= \rho_{\p},\,\,\,\,\,\Tr(\rho_{\p})=1\ee
%\be \text{Tr}(\rho_{\p})=1.\ee
By using the decomposition 
\be \rho_{\p} =Y_0\sigma_0  + \vec{Y}\vec{\sigma}\ee
we find that
\be \vec{Y}\cdot\vec{Y}=\frac{1}{4}\ee
and $Y_0 =\frac{1}{2}$. Thus the space of rays is diffeomorphic with the manifold $S^2\subset \R^3$. %(\emph{or $S^2\subset \R^4\cong u^*(2)$ with its center in $(Y_0, \vec{Y})=(\frac{1}{2},0,0,0)$? }). 
This manifold is a Hilbert manifold.\\ 
Another parametrization of the ray space could be given in terms of homogenous coordinates, say  
\be \xi = \frac{z_1}{z_2}. \ee
This description, unlike the previous one, is singular when $z_2=0$, then one may use 
\be \eta = \frac{z_2}{z_1}. \ee
The Schr\"odinger equation
\be i \hbar \frac{d}{dt} \p = H \p\ee 
on $\C^2$, which may be written in complex cartesian coordinates as
\be i \hbar \frac{d}{dt} \left(\begin{array}{c}z_1 \\z_2\end{array}\right) = \left(\begin{array}{cc}H_{11} & H_{12} \\H_{21} & H_{22}\end{array}\right) \left(\begin{array}{c}z_1 \\z_2\end{array}\right),\ee  
defines on these homogenous coordinates
a Riccati-type equation \cite{Chaturvedi:2007}
\be i \hbar \frac{d}{dt} \xi  = - H_{12}+(H_{11}-H_{22})\xi -H_{21}\xi^2.\ee  
This equation is non-linear and does not define a one parameter group of diffeomorphisms because of the singularity introduced by $z_2=0$. It is clear, however, that this behavior is an artifact of the coordinate system. It is not a singularity of the equation which describes a well defined one-parameter group of diffeomorphisms on the sphere $S^2$.  Moreover, this equation allows to remark that now the superposition rule, available on $\C^2$, is here replaced by the superposition of solutions of the Riccati equation \cite{Carinena:2001sn}. Thus the description of interference phenomena is also possible, as it should, on the manifold of pure states if we use a generalized superposition rule \cite{Man'ko:2002ti}. We shall consider now in more general terms, which tensor fields and which associated binary, bilinear brackets, are available on the Hilbert manifold of rays, pure states.

\section{Tensorial structures on pure states}

As we have stressed in the previous section, the probabilistic interpretation requires that states are identified with rays of the Hilbert space $\H$.  We have identified a description in terms of normalized states, $\braket{\p}{\p}=1$, by setting $\rho = \ket{\p}\bra{\p}$. We identify now tensors, built out of the Hermitian tensor on $\H$, which are defined on $\mathcal{R}(\H)$, i.e., they depend on complex rays rather then on states. Thus $G$ and $\Lambda$ of our previous section are modified to construct
%\be \widetilde{G}(\p)=\braket{\p}{\p}G-\braket{\p\frac{\partial}{\partial \p}}{\p\frac{\partial}{\partial \p},}\ee
\be \widetilde{G}(\p)=\braket{\p}{\p}G-(\Delta\otimes \Delta+\Gamma\otimes \Gamma)\ee
\be \widetilde{\Lambda}(\p)=\braket{\p}{\p}\Lambda-(\Delta\otimes \Gamma-\Gamma\otimes \Delta),\ee
where $\Delta$ and $\Gamma$ denote generating vector fields of $\R^+_0$-dilatations and $U(1)$-phase transformations \cite{Chruscinski:2008}. Spelled out in coordinates we find 
\begin{eqnarray}
\widetilde{G}(z,\bar{z})=&\bigg(\sum_k \bar{z}_k z_k\bigg)\sum_j \bigg(\frac{\partial}{\partial x^j}\otimes\frac{\partial}{\partial x^j}+\frac{\partial}{\partial y^j}\otimes\frac{\partial}{\partial y^j}\bigg)\nonumber\\
& -\sum_{l,m}\Bigg[ \bigg(y^l\frac{\partial}{\partial y^l}\otimes y^m\frac{\partial}{\partial y^m}\bigg)\nonumber\\
&+\bigg(y^l\frac{\partial}{\partial x^l}-x^l\frac{\partial}{\partial y^l} \bigg)
\otimes \bigg(y^m\frac{\partial}{\partial x^m}- x^m\frac{\partial}{\partial y^m} \bigg)\Bigg]
\end{eqnarray}
Similary out of $\Lambda$, we define
\begin{eqnarray}
\widetilde{\Lambda}(z,\bar{z})=&\bigg(\sum_k \bar{z}_k z_k\bigg)\bigg(\frac{\partial}{\partial y_j}\otimes\frac{\partial}{\partial x^j}-\frac{\partial}{\partial x_j}\otimes\frac{\partial}{\partial y^j}\bigg)\nonumber\\
 &-\sum_{l,m}\Bigg[\bigg(y^l\frac{\partial}{\partial y^l}+ x^l\frac{\partial}{\partial x^l} \bigg)\otimes 
\bigg(y^m\frac{\partial}{\partial x^m}- x^m\frac{\partial}{\partial y^m} \bigg)\nonumber\\
&-\bigg(y^m\frac{\partial}{\partial x^m}- x^m\frac{\partial}{\partial y^m} \bigg)\otimes
\bigg(y^l\frac{\partial}{\partial y^l}+ x^l\frac{\partial}{\partial x^l} \bigg)\Bigg]
\end{eqnarray}
By construction the bi-differential operators have the property that
\be ((f, g))^{\sim} = (\widetilde{G}+i\widetilde{\Lambda}) (df, dg)\ee
is well defined on rays whenever f and g depend only on rays and not on the representation vectors.
Thus $\widetilde{\Lambda}$ induces a Poisson bracket and $\widetilde{G}$ induces a binary, bilinear symmetric bracket on pulled-back functions from the manifold $\mathcal{R}(\H)$.\\
On functions defined on $\mathcal{R}(\H)$, which enjoy the property that their Hamiltonian vector fields are also Killing vectors for $\widetilde{G}$ when they are pulled back to $\H$, we are able again to define a $\C^*$-algebra. 
Thanks to the way we have defined our tensor fields, it turns out that the function $c(\p)=\braket{\p}{\p}$ is a central element with respect to the bracket defined by $\widetilde{\Lambda}$ on $\mathcal{F}(\H)$. It may be instructive to compute these brackets on expectation value functions of Hermitian operators, say 
\be e_A(\p)= \frac{\bra{\p}A\ket{\p}}{\braket{\p}{\p}},\,\,\,\,\,\,\,e_B(\p)= \frac{\bra{\p}B\ket{\p}}{\braket{\p}{\p}}.\ee
We find
\be \widetilde{\Lambda}(de_A, de_B)(\p) = e_{i[A, B]}(\p)\ee
while 
\be \widetilde{G}(de_A, de_B)(\p) = \2 e_{(AB+BA)}(\p)-e_A(\p)e_B(\p).\ee 
Clearly, when we consider $B\equiv \mathds{1}$, we find 
\be e_{\mathds{1}}(\p)=\frac{\bra{\p}\mathds{1}\ket{\p}}{\braket{\p}{\p}}=1\ee
and we obtain
\be \widetilde{\Lambda}(de_A, de_{\mathds{1}}) = \widetilde{G}(de_A, de_{\mathds{1}}) = 0.\ee
By using again only functions which are infinitesimal generators of unitary diffeomorphisms on $\mathcal{R}(\H)$, we may define a $\C^*$-algebra on $\mathcal{R}(\H)$ by setting 
\be ((e_A, e_B)) =  (\widetilde{G}+i\widetilde{\Lambda})(de_A, de_B) + e_A\cdot e_B.\ee
By $e_A\cdot e_B$ we mean the point wise product of the two functions. The notation $e_A, e_B$ is  reminiscent of the expectation value functions associated with Hermitian operators but this time they are not defined out of Hermitian operators. They are identified by simply requiring that their Hamiltonian vector fields are also Killing vector fields for $\widetilde{G}$. When $e_A$ and $e_B$ are extended to complex valued functions, previous requirement should be made separately for the real part and the imaginary part.\\\\
\emph{Remark:} The emerging picture of our Hilbert manifold description is that the $\C^*$-algebra approach appears to be more general then the Hilbert space approach. Indeed, to describe the formalism on the space of rays we have to go from the Hilbert space to the Hilbert manifold while we are not obliged to change perspective within the $\C^*$-algebra approach once the imaginary elements of the algebra are identified as infinitesimal generators of unitary diffeomorphisms.

\section{The GNS construction}

The $\C^*$-algebra we have defined on $\A\subset \mathcal{F}(\mathcal{R}(\H))$ allows to go from pure states to general density states by simply using the usual notion of positive, normalized linear functionals
\be \rho \in \mathcal{D}\subset \A^*.\ee
As usual with duality, from the action $\A\times \A \rightarrow \A$ of $\A$ on itself, say on the right or on the left, it is possible to induce an action on states. Starting with a state $\rho$, the action of $\A$ on $\rho$ defines a Hilbert space $\H_{\rho}$ with an Hermitian inner product
\be \braket{a}{b}_{\rho} = \rho(a^{\dagger}b),\ee
by setting 
\be (a\cdot \rho)+(b\cdot \rho) = (a+b)\cdot \rho.\ee
The set of elements which annihilate $\rho$ defines the bilateral Gelfand ideal $\mathcal{J}_{\rho}$ and 
\be \A\cdot \rho\equiv \H_{\rho}\ee
becomes identified with the quotient of $\A$ by the Gelfand ideal $\mathcal{J}_{\rho}$. When we restrict the action of elements of $\A$ on $\rho$ only by means of invertible elements, we obtain a manifold
\be \mathcal{G}\cdot \rho = \mathcal{G}\slash \mathcal{G}_{\rho} \ee
which is the quotient of the group $\mathcal{G}$ defined by the invertible elements by those which leave $\rho$ invariant. If we restrict further the invertible elements to those which preserve the pairing $\rho(e_A^{\dagger} e_B)$, the associated group becomes the `unitary group' and the corresponding manifold becomes a quotient of unitary groups.\\
The manifold picture emerging from (adjoint) group actions
%\be \phi(g)\cdot\rho=\rho_g\ee
\be g\rho g^{\dagger}=\rho_g\ee
%\be U(g)\rho U(g)^{\dagger}=\rho_g\ee
%associated to the unitary representation
%\be U: \mathcal{G}\rightarrow U(\H_{\rho}) \ee
on states $\rho$ is very interesting and clearly provides a generalization of the idea underneath coherent states. It is sufficient that out of the states generated by the group action we are able to construct a partition of unity, i.e., a completeness relation
\be \int_{\mathcal{G}\slash \mathcal{G}_{\rho}} \rho_g dg=\int_{\mathcal{G}\slash \mathcal{G}_{\rho}} \ket{\p_g}dg\bra{\p_g}=\mathds{1}.\label{cr}\ee
This aspect is closely related to the notion of tomographic set in the description of quantum mechanics by means of quantum tomography \cite{Ibort:2009}.\\
The tensorial (algebraic) structures available on the space of functions which define a $\C^*$-algebra, by duality, can be induced on the space of states. From the space of all states one may restrict them to the manifold of states selected by the action of a group or by any other means. Therefore, in suitable conditions, we may study particular problems by considering finite dimensional  real manifolds of states 
\be \mathcal{G}\slash \mathcal{G}_{\rho}.\ee
rather than the full space of states, which is usually infinite dimensional and realized by means of 
\be  L^2\big(\mathcal{G}\slash \mathcal{G}_{\rho}\big).\ee
This Hilbert space may be considered as subspace of $L^2(\mathcal{G})$  when  $\mathcal{G}_{\rho}$ is compact.
We stress again that due to the completeness relation (\ref{cr}), it is in fact possible to consider finite dimensional real submanifolds of quantum states to generate the full (infinite dimensional) space of states.\\ 
In this setting we are going to consider manifolds of states and the induced tensor fields. 

\section{Induced tensor fields on manifolds of quantum states}

On a given Hillbert manifold $\H$ we consider a covariant Hermitian tensor field 
\be \frac{\braket{d\p}{d\p}}{\braket{\psi}{\psi}}-\frac{\braket{\p}{d\p}\braket{d\p}{\p}}{\braket{\psi}{\psi}^2},\ee
admitting the property of having the generating vector field of $\C_0$-transformations in its kernel. For a given embedding of a manifold $Q$ of quantum states                                   
\be \iota_Q: Q \hookrightarrow \H,\ee
we find an induced covariant rank-2 tensor on $Q$ defined by the pull-back tensor
\be \iota^*_Q\Bigg(\frac{\braket{d\p}{d\p}}{\braket{\psi}{\psi}}-\frac{\braket{\p}{d\p}\braket{d\p}{\p}}{\braket{\psi}{\psi}^2}\Bigg).\ee
In the case that $Q$ is a homogeneous space $\mathcal{G}\slash \mathcal{G}_{0}$, we may find an embedding by means of the unitary representation
\be  \mathcal{G}\rightarrow U(\H)\label{U-repr}\ee
of a `classical' Lie group $\mathcal{G}$ on a normalized fiducial quantum state 
\be \ket{0}\in L^2(Q)\cap C^{\infty}(Q). \ee
By introducing a basis $\{\theta_j\}_{j\in J}$ of left-invariant 1-forms on $\mathcal{G}$ and a basis $\{X_j\}_{j\in J}$ on the Lie algebra of $\mathcal{G}$, we find
\be (\bra{0}R(X_j)R(X_k)\ket{0}-\bra{0}R(X_j)\ket{0}\bra{0}R(X_k)\ket{0})\theta^j\otimes\theta^k,\ee  
as pull-back tensor on $\mathcal{G}\slash \mathcal{G}_{0}$, where $\{R(X_j)\}_{j\in J}$ define the action of the Lie algebra of $\mathcal{G}$ by means of self-adjoint operators \cite{Aniello:08}. To evaluate the tensor on $T_qQ \subset u(\H)$ we set 
$\rho_0:=\ket{0}\bra{0}\in u^*(\H)$ where one finds
\begin{eqnarray}
      &  (\bra{0}R(X_j)R(X_k)\ket{0}-\bra{0}R(X_j)\ket{0}\bra{0}R(X_k)\ket{0})\theta^j\otimes\theta^k\nonumber\\
  = & (\ev{R(X_j)R(X_k)}_{\rho_0}- \ev{R(X_j)}_{\rho_0}\ev{R(X_k)}_{\rho_0})\theta^j\otimes\theta^k\\
 \equiv & T_{jk}^{\rho_0}\theta^j\otimes\theta^k.
\label{PB}\end{eqnarray}
We see that the tensor coefficients $T_{jk}^{\rho_0}$ are only dependent on the fiducial state $\rho_0$ and the chosen Lie algebra representation, but not on the individual points $q\in Q$. This coefficients decompose into a symmetric and an antisymmetric part 
\begin{eqnarray}
T_{(jk)}^{\rho_0}:= & (\ev{[R(X_j),R(X_k)]_+}_{\rho_0}- \ev{R(X_j)}_{\rho_0}\ev{R(X_k)}_{\rho_0})
\\
T_{[jk]}^{\rho_0}:= & (\ev{[R(X_j),R(X_k)]_-}_{\rho_0},
\end{eqnarray}
which allows to identify a symmetric and an anti-symmetric tensor
\begin{eqnarray}
G_{\mathcal{G}} \equiv & \,(\ev{[R(X_j),R(X_k)]_+}_{\rho_0}- \ev{R(X_j)}_{\rho_0}\ev{R(X_k)}_{\rho_0})\theta^j\odot\theta^k
\label{PBT1}\\
\Lambda_{\mathcal{G}} \equiv & \,\ev{[R(X_j),R(X_k)]_-}_{\rho_0}\theta^j\wedge\theta^k
\label{PBT2}\end{eqnarray}
on  $\mathcal{G}$.
We point out that these coefficients may be expressed by means of the expectation values
\be e_{R(X_j)}:=\frac{\bra{0}R(X_j)\ket{0}}{\braket{0}{0}}\ee
associated with the operators $R(X_j)$, i.e.\,$\,T_{(jk)}^{\rho_0}$ is  $\tilde{G}(de_{R(X_j)}, de_{R(X_k)})$ evaluated at $\rho_0$, and similarly, $T_{[jk]}^{\rho_0}$ is  $\tilde{\Lambda}(de_{R(X_j)},de_{R(X_k)})$ evaluated at $\rho_0$. In conclusion, the symmetric and the anti-symmetric pull-back structures define functions
\be g\times g \times {\mathcal{R}(\H)}\rightarrow  \R\ee
which become bi-linear on the Lie algebra $g$ once the fiducial state $\rho_0$ is fixed.\\\\
\emph{Remark:} By pulling-back the expectation-value-functions from $\H_0$ to the group we could try to define contravariant tensors on $\mathcal{G}$ by using the coefficients $T_{jk}$, however these tensors would not be defined in the directions of the isotropy group.\\\\
In general we may use any state, say a positive normalized functional $\rho \in u^*(\H)$ and consider, in analogy with (\ref{PBT1}) and (\ref{PBT2}), the classical tensors on the
group manifold associated with the quantum density state $\rho$ 
in terns of a symmetric and an imaginary skew-symmetric (0,2)-tensor
\begin{eqnarray}
L_{(jk)}^{\rho}:= & \rho ([R(X_j),R(X_k)]_+)\theta^j \odot\theta^k,\label{L-s}
\\
L_{[jk]}^{\rho}:= & \rho ([R(X_j),R(X_k)]_-)\theta^j \land\theta^k,\label{L-a}
\end{eqnarray}
respectively \cite{Aniello:09}. In the following section we will see that in particular the symmetric tensor (\ref{L-s}) will admit a direct application on the characterization of entanglement of mixed bi-partite systems.\\\\
\emph{Remark:}
One may ask whether there exits an `isometrical' embedding into a `surrounding' Hilbert space such that the later tensors (\ref{L-s}) and (\ref{L-a}) can be identified as pulled-back tensors like in the case of (\ref{PBT1}) and (\ref{PBT2}). Here indeed, by proceeding towards the more general case of manifolds of quantum operations (e.g. unitarily related Hermitian matrices \cite{Ercolessi:2001te}), we may take the intrinsic, rather the extrinsic geometric point of view. For instance, by starting from the operator-valued (0,2)-tensor field $dU^{\dagger}(q)\otimes dU(q)$ we
could define directly  
\be \rho(dU^{\dagger}(q)\otimes dU(q)),\ee
associating a left invariant tensor field on the group manifold without a dependence of an embedding.  Similarly to what happens in the GNS construction, this tensor will not be
not-degenerate. It will be degenerate along the intersection of $R(T_eQ)$ with
the Gelfand ideal associated with $\rho$. Therefore the tensor is not degenerate
on the quotient space $Q/Q_{\rho}$,  $Q_{\rho}$ being the group associated with the
sub-algebra of the Gelfand ideal.

\section{Tensor characterization of quantum entanglement}
\subsection{Pure states}
As a particular application we shall consider the problem of separability and entanglement for states of composite systems   
\be \H_A\otimes \H_B \cong \C^N \otimes \C^N.\ee
By identifying $\mathcal{G}= U(\H_A)\times U(\H_B)$ with the subgroup of transformations, which leave entanglement, resp. the Schmidt coefficients of a state invariant, we find a pull-back tensor (\ref{PB}) on 
\be U(\H_A)\times U(\H_B)\slash \mathcal{G}_{\rho_0}\ee  
which admits in the real part a Riemannian coefficient matrix
\be (G_{(jk)}) = \left(\begin{array}{cc}G^{A} & G^{AB} \\G^{AB} & G^{B}\end{array}\right)  \ee 
with the sub-block-matrices  
\begin{equation}
\label{cases}
 G_{(jk)}\equiv
\begin{cases}
G^{A}_{(jk)}=\ev{[\sigma_j,\sigma_k]_+\otimes \mathds{1}}_{\rho_0}-\ev{\sigma_j\otimes \mathds{1}}_{\rho_0}\ev{\sigma_k\otimes \mathds{1}}_{\rho_0}\\
G^{B}_{(jk)}= \ev{\mathds{1}\otimes[\sigma_j,\sigma_k]_+}_{\rho_0}-\ev{\mathds{1}\otimes\sigma_j}_{\rho_0}\ev{\mathds{1}\otimes\sigma_k}_{\rho_0}\\
G^{AB}_{(jk)}= \ev{\sigma_j\otimes \sigma_k}_{\rho_0}-\ev{\sigma_j\otimes \mathds{1}}_{\rho_0}\ev{\mathds{1}\otimes\sigma_k}_{\rho_0}
\end{cases}
\end{equation}
whenever one chooses a tensor-product representation of $U(\H_A)\times U(\H_B)$ and its Lie algebra  \cite{Aniello:09}. With this Riemannian pull-back tensor we are able to characterize entanglement without the need of performing the computational effort of a singular value decomposition into Schmidt coefficients: In particular one finds 
\begin{eqnarray}
\mbox{$\rho_0$ is separable}
\Leftrightarrow  G^{AB}=0
\Leftrightarrow  G= G^A \oplus G^B.
\end{eqnarray}
Moreover, it turns out the the sub-block-matrix $G^{AB}$ is useful to the compute the \emph{distance to separable states}, which has been identified in \cite{Man'ko:2002ti} by the quantity
\be \Tr((R)^{\dagger}R)\ee
with 
\be R:= \rho_0 - \rho^A_0\otimes \rho^B_0.\ee
Here we find
\be \Tr((G^{AB})^{T}G^{AB}) \sim \Tr((R)^{\dagger}R).\ee
On the other hand we recall that the pull-back tensor offers in its imaginary part also an anti-symmetric tensor. In particular on the same homogeneous space of entangled quantum states we have therefore not only Riemannian, but also pre-symplectic tensor coefficients to evaluate. Here we encounter the coefficient matrix
\be (\Lambda_{(jk)}) = \left(\begin{array}{cc}\Lambda^{A} & 0 \\0 & \Lambda^{B}\end{array}\right)  \ee 
with 
\begin{equation}
\label{cases}
 \Lambda_{[jk]}\equiv
\begin{cases}
 \Lambda_{[jk]}^{A} =\ev{[\sigma_j,\sigma_k]_-\otimes \mathds{1}}_{\rho_0}\\
\Lambda_{[jk]}^{B}= \ev{\mathds{1}\otimes[\sigma_j,\sigma_k]_-}_{\rho_0}\\
 0.\\
\end{cases}
\end{equation}
For what concerns the entanglement of the fiducial state, it turns out that the pre-symplectic part of the pull-back tensor behaves in an opposite way to the Riemannian part \cite{Aniello:09}: 
\begin{eqnarray}
\mbox{$\rho_0 $ is maximal entangled}
\Leftrightarrow \Lambda= \Lambda^A = \Lambda^B=0.
\end{eqnarray}

\subsection{Mixed states}

We may end up with an identification of an applicable tensor constructions for the entanglement characterization of mixed states. The latter are identified in the Schr\"odinger picture as elements of the convex hull 
\be D(\H_A\otimes \H_B)\cong D(\C^N\otimes \C^N)\ee
of pure states associated to the bi-partite system $\H=\C^N\otimes \C^N$ discussed in the previous section.  In particular, by evaluating the coefficients $L_{(jk)}$ of the intrinsic geometrically defined rank-2 tensor field (\ref{L-s}) on the orbits generated by the local unitary group $U(N)\times U(N)$ in $u^*(\H)$, we find a neat connection to a separability criteria 
proposed by de Vicente \cite{deVicente:2007}. By analogy to the case of pure states orbits we consider $\rho\in D(\C^N\otimes\C^N)$, a density state on an orbit $Q/Q_{\rho}$ generated by $Q=U(N)\times U(N)$. The tensor coefficients
\be L_{(jk)}= \Tr(\rho [R(X_j), R(X_k)]_+ ),\label{L}\ee  
defined on this orbit may then become identified according to the coefficient matrix
\be (L_{(jk)}) = \left(\begin{array}{cc}L^{A} & L^{AB} \\L^{AB} & L^{B}\end{array}\right)  \ee 
with the sub-block-matrices 
\begin{equation}
\label{cases}
L_{(jk)}\equiv
\begin{cases}
L^{A}_{(jk)}=\ev{[\sigma_j,\sigma_k]_+\otimes \mathds{1}}_{\rho}\\
L^{B}_{(jk)}= \ev{\mathds{1}\otimes[\sigma_j,\sigma_k]_+}_{\rho}\\
L^{AB}_{(jk)}= \ev{[\sigma_j\otimes \mathds{1},\mathds{1}\otimes  \sigma_{k}]_+}_{\rho}
\end{cases}
\end{equation}
This intrinsic geometrically defined tensor admits now the following application:
\begin{eqnarray}
\mbox{$\rho_0$ is separable}
\Rightarrow  \frac{2}{N}\Tr(\sqrt{(L^{AB})^{\dagger}L^{AB}}) \leqslant \2(N^2-N)\label{ineq}\\\nonumber\\
\Rightarrow \mbox{$\rho_0$ is separable (for N=2).}\label{de Vicente 1}
\end{eqnarray}
One notes that the required coefficients are given in the off-diagonal block elements $L_{(jk)}^{AB}:=C_{(jk)}$ of the coefficient matrix $(L_{(jk)})$, defined by
\be C_{(jk)}= \Tr(\rho [\sigma_j\otimes \mathds{1},\mathds{1}\otimes  \sigma_{k-N^2}]_+),\ee
yielding
\be C_{(jk)}=\Tr(\rho \sigma_j\otimes \sigma_{k-N^2}).\ee
One completes then directly the proof by means of Theorem 1 and following discussion leading to Corollary 1 in \cite{de Vicente:2007}, by identifying the Ky Fan Norm  
\be \|C\|_{KF} :=\Tr(\sqrt{C^{\dagger}C}),\ee
within the inequality (\ref{ineq}). 
Let us apply the above criterion on an explicit example:
\\\\
{\bf Example: Werner states for the case $N=2$}
\\\\
Consider a density state in $D(\C^2\otimes \C^2)$, defined as convex combination of a maximal entangled pure state
\be \ket{\phi^+}:=\frac{1}{\sqrt{2}}\bigg(\ket{0}\otimes \ket{0}
 + \ket{1}\otimes \ket{1}\bigg),\ee
and a maximal mixed state \be\rho^*:=\frac{1}{4}\mathds{1},\ee
according to  
\be \rho_W:=x\ket{\phi^+}\bra{\phi^+}+(1-x)\rho^*\ee
with $x\in [0,1]$. The latter state is referred in the literature to the class of Werner states \cite{Werner:1989zz}. By identifying 
\be  \rho_W \equiv \rho \in \A^*\cong M_4(\C),\ee
we find a symmetric tensor on a 6-dimensional real submanifold \be SU(2)\times SU(2)\subset U(4)\subset\A^*\ee 
whose coefficient matrix reads 
\be (L_{(jk)})= \left(
\begin{array}{llllll}
 1 & 0 & 0 & x & 0 & 0 \\
 0 & 1 & 0 & 0 & -x & 0 \\
 0 & 0 & 1 & 0 & 0 & x \\
 x & 0 & 0 & 1 & 0 & 0 \\
 0 & -x & 0 & 0 & 1 & 0 \\
 0 & 0 & x & 0 & 0 & 1
\end{array}
\right), \ee
where one identifies within a decomposition 
\be (L_{(jk)}):=\left(\begin{array}{cc}A & C \\C & B\end{array}\right),\ee
the block of diagonal elements
\be C = \left(
\begin{array}{lll}
 x & 0 & 0 \\
 0 & -x & 0 \\
 0 & 0 & x
\end{array}
\right). \ee  
The latter is identical related to the symmetric tensor coefficients $L_{(jk)}$ for $1 \le j \le 3$ and
 $5 \le k \le 6$.  By computing the Ky Fan Norm of $C$ one finds
\be \Tr(\sqrt{C^{\dagger}C}) = 3 x,\ee
where we conclude according to the criterion  (\ref{de Vicente 1}) that $\rho_W$ is separable iff
\be x \leqslant \frac{1}{3}.\label{1/3}\ee

\section{Conclusions and outlook}

We have shown that a geometrical description of quantum mechanics is possible and that many concepts and constructions available in classical mechanics are also available in the quantum frame work. The richer structure emerging in the quantum setting allows to introduce not only Poisson brackets but also Jordan brackets and Lie-Jordan algebras, a description based just on observables as advocated by Segal. We have also shown that separability and entanglement of quantum states for composite systems may be given a geometrical description by means of symmetric and skew-symmetric tensors. Clearly, the consideration of Hamiltonian vector fields allows to consider transformations and evolution on the manifold of states in a context of `classical' differential geometry, making therefore available all the experience acquired in the classical setting.\\
Most of our constructions rely on tensor fields with values in the tensor algebra of self-adjoint operators 
\be dU^{\dagger}(g)\otimes dU(g) = -(R(X_j)R(X_k))\theta^j\otimes \theta^k.\ee    
Inspired by classical mechanics where Poincar\'e has introduced anti-symmetrized products of the symplectic structure and of the symplectic potential  (absolute Poincar\'e invariants) one may introduce higher order tensors by means of similar constructions in terms of 
\be (dU^{\dagger}(g)\otimes dU(g))\otimes(dU^{\dagger}(g)\otimes dU(g))\otimes..\otimes(dU^{\dagger}(g)\otimes dU(g))\ee 
As we will show elsewhere, this approach allows to deal with n-fold expectation values and correlation functions. We believe that a worked out geometrization of quantum theories may provide very useful suggestions for a unification of quantum mechanics and general relativity.

\subsection*{Acknowledgments}
This work was supported by the National Institute of Nuclear Physics (INFN).\\
G. F. V. would like to thank G. Marmo for kind hospitality.


\begin{thebibliography}{35}


% \cite{Ercolesi:2007}
\bibitem{Ercolesi:2007}
Ercolessi E, Ibort L A,  Marmo G and  Morandi G 2007 %Alternative Linear Structures for Classical and Quantum Systems,
\emph{Int. J. Mod.\ Phys. A}  {\bf 22} 3039

%\cite{Perelomov:1971bd}
\bibitem{Perelomov:1971bd}
Perelomov A M 1972 %Coherent states for arbitrary Lie groups, 
\emph{Commun.\ Math.\ Phys.}  {\bf 26} 222
  %%CITATION = CMPHA,26,222;%%

%\cite{Gilmore:1972}
\bibitem{Gilmore:1972}
Arecchi F T, Courtens E, Gilmore R and Thomas H 1972 %Atomic Coherent States in Quantum Optics,
\emph{Phys. Rev. A} {\bf 6} 2211--2237

%\cite{Onofri:1974}  
\bibitem{Onofri:1974}
Onofri E 1975 %A note on coherent state representations on Lie groups,
\emph{Journ. Math. Phys.} {\bf 16} 1087

%\cite{Man'ko:1996xv}  
\bibitem{Man'ko:1996xv}
Man'ko V I, Marmo G, Sudarshan E C G and Zaccaria F 1997 %f-Oscillators and Nonlinear Coherent States,
\emph{Phys.\ Scripta} {\bf 55} 528

%\cite{Heslot:1985}    
\bibitem{Heslot:1985}
Heslot A 1985 %Quantum mechanics as a classical theory, 
\emph{Phys. Rev. D} {\bf 31} 1341--1348 

\bibitem{Rowe:1980}
Rowe D J, Ryman A and Rosensteel G 1980 %Many body quantum mechanics as a symplectic dynamical system, 
\emph{Phys Rev A} {\bf 22} 2362--2372 

\bibitem{Cantoni:1975}
Cantoni V 1975 %Generalized transition probability, 
\emph{Comm. Math. Phys.} {\bf 44} 125--128 

\bibitem{Cantoni:1977a}
Cantoni V 1977 %The Riemannian structure on the space of quantum-like systems, 
\emph{Comm. Math. Phys.} {\bf 56} (1977),  189--193  

\bibitem{Cantoni:1977b}
Cantoni V 1977 %Intrinsic geometry of the quantum-mechanical phase space, Hamiltonian systems and Correspondence Principle, 
\emph{Rend. Accad. Naz. Lincei} {\bf 62} 628--636
 
\bibitem{Cantoni:1980} 
Cantoni V 1980 %Geometric aspects of Quantum Systems, 
\emph{Rend. sem. Mat. Fis. Milano} {\bf 48} 35--42

\bibitem{Cantoni:1985}
Cantoni V 1985 %Superposition of physical states: a metric viewpoint, 
\emph{Helv. Phys. Acta} {\bf 58} 956--968 
 
\bibitem{Cirelli:1983} 
Cirelli R, Lanzavecchia P and Mani\`a A 1983 %Normal pure states of the von 
%Neumann algebra of bounded operator as K\"ahler manifold, 
\emph{J. Phys. A: Math. Gen.} {\bf 16} 3829--3835  

\bibitem{Cirelli:1984}
Cirelli R and Lanzavecchia P 1984 %Hamiltonian vector fields in Quantum Mechanics, 
\emph{Nuovo Cimento B} {\bf 79} 271--283

\bibitem{Abbati:1984}
Abbati M C, Cirelli R, Lanzavecchia P and Mani\`a A 1984 %Pure states of 
%general quantum mechanical systems as K\"ahler bundle, 
\emph{Nuovo Cimento B} {\bf 83} 43--60  

\bibitem{Ashtekar:1997ud}
Ashtekar A and Shilling T A  1999 {\it Geometrical Formulation of Quantum Mechanics} (On Einstein's Path: Essays in Honor of Engelbert Sch\"ucking, Proceedings 
of a symposium held at the Physics department in New York University) ed A Harvey  (New York: Springer) 23--65 

\bibitem{Gibbson:1992}
Gibbson G W 1992 %Typical states and density matrices, 
{\it J. Geom. Phys.} {\bf 8} 147--162  

\bibitem{Brody:2001}
Brody D and Hughston L P 2001 %Geometric quantum mechanics, 
{\it J. Geom. Phys.} {\bf 38} 19--53  

%\cite{deGosson:2007}
\bibitem{deGosson:2007}
de Gosson M 2001 {\it The principles of Newtonian and quantum mechanics} (London:Imperial College Press)

%\cite{Man'ko:2005:gqm}
\bibitem{Man'ko:2005:gqm}
Man'ko V I, Marmo G, Sudarshan E C G and Zaccaria F 2005 %The geometry of density states 
{\it Rep. Math. Phys. } {\bf 55} 405--22

%\cite{Carinena:2007:gqm}
\bibitem{Carinena:2007:gqm}
Carinena J F, Clemente-Gallardo J and Marmo G 2007  %Geometrization of quantum mechanics
{\it Theoretical and Mathematical Physics }{\bf 152} 894--903 
 

%\cite{Clemente:2008}
\bibitem{Clemente:2008}
Clemente-Gallardo J and Marmo G 2008 %Basics of Quantum Mechanics, Geometrization and some Applications to Quantum Information,
\emph{Int. J. Geom. Meth. Modern Physics} {\bf 5} 989--1032
 
%\cite{Chernoff:1974}
\bibitem{Chernoff:1974}
Chernoff P and Marsden J E  1974 {\it Properties of Infinite Dimensional Hamiltonian Systems} %{\it Lecture Notes in Math.} {\bf 425}, 
(New York: Springer)  

%\cite{Schmid:1987} 
\bibitem{Schmid:1987}
Schmid R 1987 {\it Infinite dimensional Hamiltonian Systems} (Naples: Bibliopolis)

%\cite{Lang:1994} 
\bibitem{Lang:1994}
Lang S 1995 {\it Differential and Riemannian Manifolds} 
(New York: Springer) %p?103
 
%\cite{Majorana:1932}

%\bibitem{Majorana:1932}
%E. Majorana, Atomi orientati in campo magnetico variabile, \emph{Nuovo Cimento} {\bf 9}, 43 (1932). 

%\cite{Carinena:2001sn}
\bibitem{Carinena:2001sn}
Carinena J F, Grabowski J and Marmo G 2001 %Some physical applications of systems of differential equations admitting a superposition rule,
\emph{Rept.\ Math.\ Phys.\ } {\bf 48} 47
  %%CITATION = RMHPB,48,47;%%

%\cite{Chaturvedi:2007}
\bibitem{Chaturvedi:2007}
Chaturvedi S, Ercolessi E,  Marmo G,  Morandi G,  Mukunda N and Simon R 2007 {\it Pramana J. Phys.} {\bf 69} 317--327


%\cite{Chruscinski:2008}
\bibitem{Chruscinski:2008}
Chruscinski D and Marmo G 2009 %Remarks on the GNS Representation and the Geometry of Quantum States, 
{\it Open Syst.Info.Dyn.}{\bf 16} 157--177

%\cite{Ibort:2009}
\bibitem{Ibort:2009}
Ibort A, Man'ko V I, Marmo G, Simoni A and Ventriglia F 2009 %An introduction to the tomographic picture of quantum mechanics,  
{\it Phys. Scr.} {\bf 79} 065013


%\cite{Aniello:08}
\bibitem{Aniello:08}
Aniello P, Marmo G, Volkert G F 2009 %Classical tensors from quantum states,
 \emph{Int.\ J.\ Geom.\ Meth.\ Mod.\ Phys.} {\bf 06} 369--383  

%\cite{Aniello:09}
\bibitem{Aniello:09}
Aniello P,  Clemente-Gallardo J, Marmo G and Volkert G F 2009 %Classical tensors and quantum entanglement I: Pure states, 
\emph{Int.\ J.\ Geom.\ Meth.\ Mod.\ Phys.} {\bf 07} 485--503 


%\cite{Ercolessi:2001te}
\bibitem{Ercolessi:2001te}
Ercolessi E, Marmo G and Morandi G 2001%Geometry of mixed states and degeneracy structure of geometric phases for multi-level quantum systems. A unitary group approach,
\emph{Int.\ J.\ Mod.\ Phys. A} {\bf 16} 5007

%\cite{Man'ko:2002ti}  
\bibitem{Man'ko:2002ti}
Man'ko V I, Marmo G, Sudarshan E C G and Zaccaria F 2002 %Interference and entanglement: an intrinsic approach, 
\emph{J.\ Phys.\ A}  {\bf 35} 7137

%\cite{de Vicente:2007}
\bibitem{deVicente:2007}
de Vicente J I 2007 %Separability criteria based on the Bloch representation of density matrices, 
\emph{Quantum Inf. Comput.} {\bf 7} 624

%\cite{Werner:1989zz}
\bibitem{Werner:1989zz}
Werner R F 1989 %Quantum states with Einstein-Podolsky-Rosen correlations admitting a hidden-variable model, 
\emph{Phys. Rev. A} {\bf 40} 4277


\end{thebibliography}
\end{document}